          [1999/12/02 1.01 (PWD)]
\documentclass[a4paper,twocolumn]{esapub} 

\def\grs{\mbox{GRS } 1915+105}

\def\saxjdhdn{\mbox{SAX J1819.3-2525}}

\def\cmmoinsdeux{\mbox{ cm}^{-2}}

\def\kms{\mbox{ km\,s}^{-1}}

\def\kpc{\mbox{ kpc}}

\def\Msol{\mbox{ }M_{\odot}}

\def\ergs{\mbox{ erg\,s}^{-1}}

\def\deg{^{\circ}}

\def\amin{^\prime}

\def\asec{^{\prime \prime}}
\def\asecp{{\rlap.}^{\prime \prime}}

\def\heu{^{h}}

\def\hmin{^{m}}

\def\hsecp{{\rlap.}^{s}}

\def\nh{N \mbox{(H)}}

\def\ltsima{\; \buildrel < \over \sim \;}
\def\simlt{\lower.5ex\hbox{\ltsima}}            
\def\gtsima{\; \buildrel > \over \sim \;}
\def\simgt{\lower.5ex\hbox{\gtsima}}            

\input{psfig}				%

\usepackage{natbib,latexsym}

\title{On the nature of the microquasar V4641 Sagittarii}
\author{S. Chaty}
\affil{Department of Physics and Astronomy, The Open University, 
United Kingdom, s.chaty@open.ac.uk}
\author{I.F Mirabel}
\affil{Service d'Astrophysique, DSM/DAPNIA/SAp, Centre d'\'Etudes de Saclay, 
France}
\affil{Instituto de Astronom\'{\i}a y F\'{\i}sica del Espacio, Conicet, 
Argentina, mirabel@discovery.saclay.cea.fr}
\author{J. Mart\'{\i}}
\affil{Departamento de F\'{\i}sica, Escuela Polit\'ecnica Superior, 
	Universidad de Ja\'en, Spain, jmarti@ujaen.es}
\author{L.F. Rodr\'{\i}guez}
\affil{Instituto de Astronom\'{\i}a, Campus UNAM, Morelia, 
	Michoac\'an 58190, M\'exico, luisfr@astrosmo.unam.mx}

\begin{document}


\maketitle

\begin{abstract}
We present photometric and spectroscopic optical and near-infrared (NIR)
observations\footnote{Based 
on observations collected at the 
European Southern Observatory, Chile (ESO ID 63.H-0493 and 64.H-0382)} 
taken during the outburst of the recently flaring source 
V4641 Sgr = SAX J1819.3-2525 \citep{in'tzand:2000}, 
on September 1999.
This source was independently detected as the {\it RXTE} source XTE J1819-254 
\citep{markwardt:1999a}, and afterwards identified with the
variable object V4641 Sgr \citep{kato:1999}.
It underwent a bright optical outburst
on 1999 Sept. 15.7 UT, going from magnitude 14
to 8.8 in the V-band  \citep{stubbings:1999}, 
and reaching 12.2 Crab in the X-rays
\citep{smith:1999} and $Ks \simeq 13$. 
This outburst was therefore bright, but very brief,
with an e-fold decay time of 0.6 days.
A radio source was resolved, making of V4641 Sgr a new microquasar 
\citep{hjellming:2000}.
We discuss the nature of this system, showing 
that our observations suggest a distance farther
than previously derived from the radio observations \citep{hjellming:2000}.
The distance of the system would be between 3 and 8 kpc, 
the companion star being a B3-A2 main sequence star. Another possibility
is that the companion star is crossing the Hertzsprung gap (type B3-A2 IV),
and in this case the distance cited above would be the minimum distance
of the system.
The system is therefore an Intermediate or 
High Mass X-ray Binary System (IMXB or HMXB).
The inconsistency regarding the distance
between the radio and optical/NIR observations
could be explained by the detection of an interaction
between matter ejected before the X-ray outburst and the surrounding
medium of the source. If this is confirmed, this source could be
added to the short list of microquasars where such an interaction
has been detected.
\end{abstract}

\section{Introduction}

The source $\saxjdhdn$ attracted considerable attention after 
the detection of a giant optical outburst on 1999 September 15.7 UT, 
from the magnitudes 14.0 to 8.8 in the V-band \citep{stubbings:1999}.
The source, in the direction of the galactic bulge,
 was centered on a star called V4641 Sgr in the
catalog of variable stars 
and located in the constellation of Sagittarius \citep{kazarovets:2000}. 
Due to a confusion, most of the references of this X-ray
source are reported to the name of GM Sgr. 
After this confusion was clarified by \citet{williams:1999} and
\citet{samus:1999},
the source was newly designated V4641 Sgr \citep{kazarovets:2000}.
The X-ray source XTE J1819-254 flared, from 1.6 to 12.2 Crab in the X-rays
on 1999, September 14th, as observed by {\it RXTE} in the 2-12 keV band, 
through a brief but dramatic eruption (brightest
source of X-rays in the sky), its position being coincident with
the optical transient \citep{smith:1999}.
Less than 10 hours later, the source was fainter than 50 mCrab.
This source was identified with the previously detected 
faint X-ray transient, $\saxjdhdn$, 
discovered by {\it SAX} on 1999, Feb 20th, 
 \citep{in'tzand:1999} (energy $0.012-0.3$ Crab), 
and independently detected by {\it RXTE}, 
on 1999, Feb 18th \citep{markwardt:1999a}, with an
energy between $<1$ and $80$ mCrab in the 2-10 keV energy band.

\cite{kato:1999} reported an unusual optical activity prior to this giant
optical and X-ray outburst, through a $\sim 1$ mag high-amplitude modulation
6 days before the giant outburst (on Sep. 8th),
and a quasi-periodicity of 2.5 days, which they claimed to 
correspond to the orbital period.
Three other eruptions
followed each lasting less than two hours in the X-rays, ones of the fastest
bursts ever seen. 
The observations by {\it RXTE} \citep{wijnands:2000} 
allowed to observe some strong flaring activity: 
fluctuations by a factor of 4 on the timescale of seconds,
and 500 on minutes. 
No QPO was detected,
but some red noise at $<$ 30 Hz was present.
The observations by {\it SAX} \citep{in'tzand:2000} 
gave a best fit for the column density of
$\nh \sim 0.05 \pm 0.02 \times 10^{22} \cmmoinsdeux$.

The VLA radio telescope detected on Sept. 16.02 UT
a strong radio source of 0.4 Jy at 4.9 GHz, at the
position of the variable star, at 
$\alpha$ = $18\heu 19\hmin 21\hsecp636$,
$\delta$ = $-25\deg 24\amin 25\asecp6$ (J2000) \citep{hjellming:1999a}. 
The galactic coordinates are ($l,b$) = ($6.774015\deg, -4.789045\deg$).
The flux decreased on a timescale of hours, 
with an e-fold decay time of 0.6 day.
The source was resolved, with the presence of an elongation
extending $0.25 \asec$ between 0.6-1.2 day after the huge X-ray flare.
On Sept. 17.93, 22.00 and 24.1 UT, the elongation was at the same position
\citep{hjellming:1999b}.
They claimed that the proper motion was $0.5\asec$ / day, 
but this is strongly
dependent upon the time of the ejection. The 
jets seem to suggest a high inclination angle. 
This allowed to classify the source as a new microquasar
(for a review on the jet sources see \cite{mirabel:1999}).
An HI absorption experiment
towards the source implied a distance $d > 0.4 \kpc$ \citep{hjellming:2000},
and these authors proposed a likely distance of $0.5 \kpc$.

\cite{orosz:2000a} derived from ESO spectroscopic observations
an optical mass function of $2.74 \pm 0.12 \Msol$, which, combined
with the information on the inclination, makes of V4641 a black hole system
with a mass of the compact object $8.73 \leq M_1 \leq 11.70 \Msol$. 
They also found a spectroscopic
period of $2.81678 \pm 0.00056$ days, and assuming an extinction
$E(B-V) = 0.32 \pm 0.10$, quoted a distance between $7.40$ and $12.31 \kpc$
(note that this is bigger than previously derived by \cite{orosz:2000}).

Through our on-going ESO Target of Opportunity program
 aimed at observing new X-ray flaring sources,
we got quickly NIR and optical imaging 
and spectroscopic observations
of this new transient source during its outburst and
we could follow it on its decline for a few months
(from 1999, September to 2000, June).
We report in Section \ref{obs} the main observations and results, 
and discuss them in Section \ref{discussion}.
A more detailed study of this source is reported in Chaty et al. (in prep.).


\section{Observations and results} \label{obs}

\begin{figure}
\centerline{\psfig{file=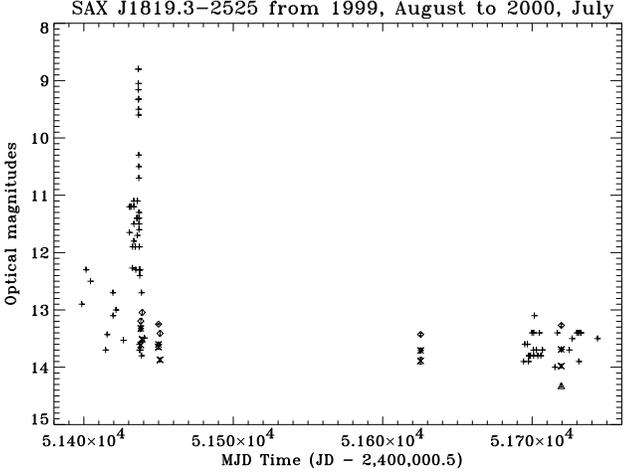,angle=+90.,width=8.9cm}}
\caption[]{Optical Observations. +:VSNET, $\triangle$: B, $\times$:V, $\ast$:R, $\diamond$:I magnitudes. 
The beginning of the optical activity took place on 1999 Sept. 8 UT
(= MJD 51429.5), followed by the outburst of 1999 Sept. 15.7 UT (= MJD 51437).
\label{optique_tout}}
\end{figure}

\begin{figure}
\centerline{\psfig{file=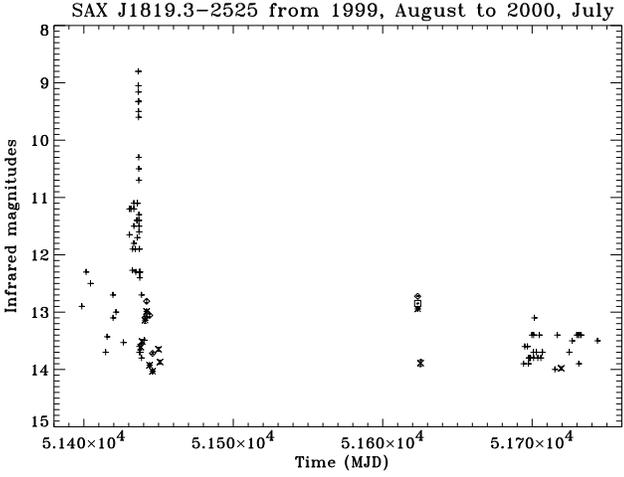,angle=+90.,width=8.9cm}}
\caption[]{NIR (+ optical) observations. 
+:VSNET, $\times$:V, $\ast$:J, $\Box$: H, 
$\diamond$:Ks magnitudes. \label{vir_tout}}
\end{figure}

\begin{figure}
\centerline{\psfig{file=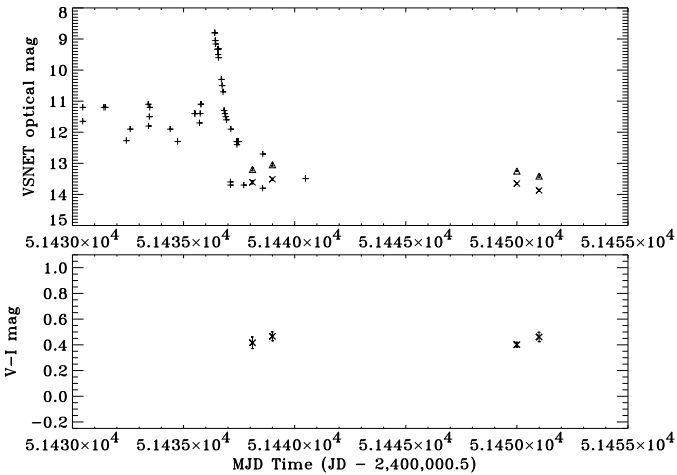,angle=+90.,width=9.cm}}
\caption[]{Top: +:VSNET, $\times$:V, $\triangle$:I magnitudes. 
Bottom: V-I color. \label{figure_V-I}}
\end{figure}

\begin{figure}
\centerline{\psfig{file=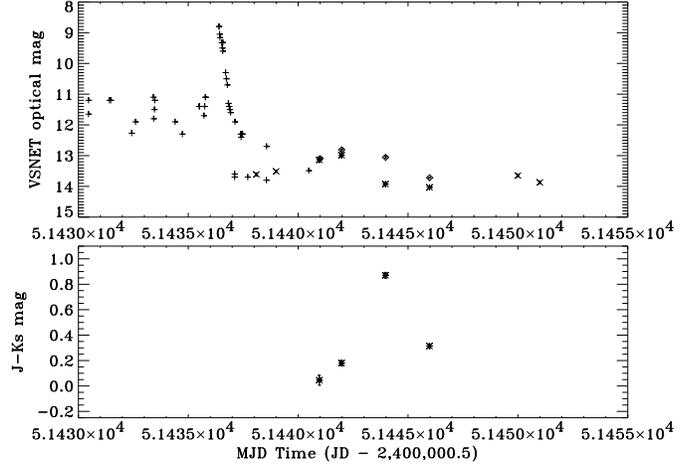,angle=+90.,width=8.9cm}}
\caption[]{Top: +:VSNET, $\times$:V, $\ast$:J, $\diamond$:Ks magnitudes). 
Bottom: J-Ks color. \label{figure_J-K}}
\end{figure}

The optical imaging and spectroscopic observations took place on
1999, September 16, 17, 28 and 29 and on 2000, March 21 and June, 24; 
and the infrared imaging and
spectroscopic observations on 1999, September 19, 20,
22 and 24, and on 2000, March 20. All the observations took
place at ESO, La Silla, but those of June, 24, which were performed 
at the 1.23 m telescope of the Centro
Astron\'omico Hispano Alem\'an at Calar-Alto, 
with the CCD optical camera and exposure times
between 30 and 60s (more details on these observations
are reported in \citet{marti:2001}).
The airmass during the observations was always between 1.008 and 1.020.
The optical observations were performed 
with the NTT telescope and the instrument EMMI RILD.
We imaged the source in V, R and I filters, and took
some spectra with the grism \#1.
The exposure times were nearly 5 min each for the imaging
and 15 min for the spectroscopy.
The infrared observations 
were performed with the NTT and the instrument SOFI.
The imaging was taken through the filters J, H and Ks,
and the spectra with the Grism Red (GR) and 1" slit.
The exposure times were chosen as 15 min for the infrared imaging
and spectroscopy.
The Tables \ref{table_observations_optique} and 
\ref{table_observations_infrarouge} summarize respectively all the different 
optical and infrared observations.
The lightcurves of the overall optical and infrared observations
are respectively reported in Figures \ref{optique_tout} and 
\ref{vir_tout}.
The V-I and J-K colors during the outburst are respectively reported 
in Figures \ref{figure_V-I} and \ref{figure_J-K}.

\begin{table*}
\begin{flushleft}
\begin{tabular}{|c|c|c|c|c|c|c|} \hline
{\em Date}      & {\em MJD}     & {\em Inst}    & 
{\em B}   & {\em V} & {\em R} & {\em I} \\ \hline


16/09/99  & 51438.1 & EMMI      
& - & $13.611\pm0.095$ & $13.325\pm0.1$ & $13.195\pm0.085$ \\
17/09/99  & 51439.0& EMMI      
& - & $13.511\pm0.06$  &       -        & $13.045\pm0.08$ \\
28/09/99  & 51450.0& EMMI             
& - & $13.649\pm0.062$   & $13.603\pm0.012$ & $13.248\pm0.016$ \\
29/09/99  & 51451.0& EMMI 
& - & $13.87\pm0.064$  & -                & $13.409\pm0.083$ \\
21/03/00  & 51625.4 & EMMI 
& - & $13.889\pm0.116$ & $13.708\pm0.075$ & $13.428\pm0.043$ \\
24/06/00 & 51719.5 & C.A.
& $14.32\pm0.05$ & $13.98\pm0.05$   & $13.69\pm0.05$   & $13.27\pm0.05$ \\

\hline
\end{tabular}
\end{flushleft}
\caption[]{\label{table_observations_optique} {\bf Diary of the 
optical observations.} C.A.: Calar Alto \\
}
\end{table*}

\begin{table*}
\begin{flushleft}
\begin{tabular}{|c|c|c|c|c|c|c|} \hline
{\em Date} & {\em MJD}& {\em Inst} & {\em J     } &  {\em H} &  {\em Ks} & {\em J-Ks}     \\ \hline


%
19/09/99  & 51441.0& SOFI      
& $13.14\pm0.113$  & -              & $13.095\pm0.04$  & 0.045 $\pm$ 0.153 \\
20/09/99  & 51441.98& SOFI 
& $12.989\pm0.026$ & -              & $12.809\pm0.056$ & 0.18  $\pm$0.082 \\
22/09/99  & 51443.97& SOFI      
& $13.927\pm0.03$    & -              & $13.056\pm0.058$ & 0.870 $\pm$ 0.088 \\
24/09/99  & 51445.98& SOFI           
& $14.033\pm0.027$   & -              & $13.719\pm0.029$ & 0.314 $\pm$0.056 \\
20/03/00  & 51623.42& SOFI             
& $12.944\pm0.007$   & $12.849\pm0.01$  & $12.723\pm0.013$ & 0.221 $\pm$0.02 \\
%
\hline
\end{tabular}
\end{flushleft}
\caption[]{\label{table_observations_infrarouge} {\bf Diary of the infrared observations.} \\
}
\end{table*}

After the big outburst (from V=14 to 8.8 mag), 
there was still some flaring activity in V, R and I
with variations of $\sim 0.5$ mag with no significant change in the colors.
In NIR there was also 
some flaring activity with variation of $\sim 1$ mag in J and K,
with a significant change in the J-K color during the post-outburst
(between 2 and 5 days after the giant outburst).
This suggests an increased K-contribution compared to J, which 
can be explained either by the emission of a jet,
or the appearance of heated dust (as seen in the case of $\grs$
by \cite{mirabel:1996a}), or even by the interaction 
with the interstellar medium.

\begin{figure}
\centerline{\psfig{file=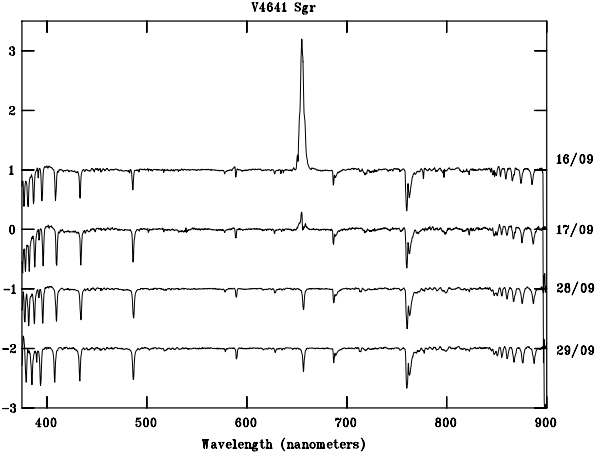,angle=+90.,width=9.cm}}
\caption[]{Normalized and offset optical spectra. \label{spec_opt_norm}}
\end{figure}

We report the normalized
optical spectra offset to get an easier reading 
in Figure \ref{spec_opt_norm}.
The first striking fact is that on a timescale of one day,
the lines were changing from emission to absorption.
All the Balmer serie is visible: H$\alpha$, $\beta$, $\gamma$, $\delta$, 
$\epsilon$, $\zeta$...
The H$\alpha$ emission line is extraordinarily strong: 
one day after the outburst, its
equivalent width was $\sim 100 \AA$, with a FWZI of 
$\sim 6700 \kms$ and a blue wing.
There was also a strong He I $5876 \AA$.
The Na-D absorption line equivalent width of $0.45 \AA$ gives E(B-V) = 0.25 
\citep{munari:1997} implying 
$\nh = 0.14 \times 10^{22} \cmmoinsdeux$ \citep{bohlin:1978}.
There is a strong variability of the lines: 
Balmer H$\alpha$, H$\beta$ and also of He I.
The He II $4680 \AA$ line was claimed to be prominent in emission 
nearer to the outburst time 
by \cite{ayani:1999}. Since we could not detect it, this line
was also very variable.
We can also note the blue continuum, visible on the flux calibrated spectra,
suggesting the emission from an accretion disk, or from a corona.

In the NIR, the He I and Br $\gamma$ lines were observed as strong
lines by \cite{charles:1999a}, in a UKIRT/CGS4 spectrum taken
on Sept 17.22 UT. 
He I exhibited an equivalent width of  2.1 nm, and Br $\gamma$ of 1.4 nm,
both showing extended blue wings (FWZI=5900 km/s),
suggesting a high velocity wind component \citep{charles:1999a}.
The continuum of our NIR spectrum was blue, like the spectrum from
\cite{charles:1999a}.
However, our NIR spectrum shows only 
very faint HeI, He II and Br$\gamma$ lines,
which appear therefore to be strongly variable.
The emission lines and the variability suggest the
 accretion of matter onto a compact
object with a high-velocity wind component ($\sim 6000 \kms$), 
also perhaps the presence of a cocoon or a jet.

\section{Discussion} \label{discussion}

\subsection{Nature of the system: the companion star}

\begin{figure}
\centerline{\psfig{file=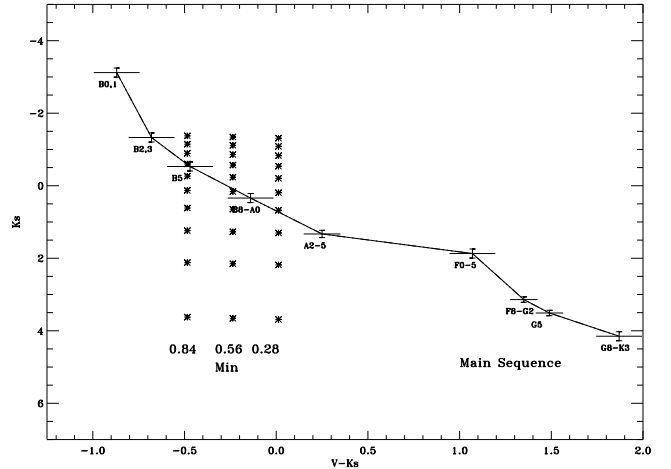,angle=+90.,width=9.cm}}
\caption[]{Color-magnitude [V-Ks,Ks] diagram. *: Min magnitudes of V4641 Sgr. 
+: typical main sequence stars \citep{ruelas-mayorga:1991}. 
0.28, 0.56 and 0.84 are the visual absorptions
corresponding respectively to the column densities 
$0.05$, $0.1$ and $0.15 \times 10^{22} \cmmoinsdeux$.
\label{hr}}
\end{figure}

We plotted on a color-magnitude diagram (CMD, Figure \ref{hr})
the optical and NIR magnitudes when the source was faint.
We plotted in this figure the absolute magnitudes corrected for 
three different values of the absorption, and corresponding to
ten different values of the distance of the source.
The values of the absorption correspond 
to $0.05$, $0.1$ and $0.15 \times 10^{22} \cmmoinsdeux$,
derived from our observations and from {\it SAX} observations 
\citep{in'tzand:2000}.
The values of the distance of the source go 
from 1 to $10 \kpc$.
If we constrain the companion object of the binary system 
to be a main sequence star, 
its location on the CMD, 
taking into account the uncertainty on the absorption,
suggests that the distance is constrained to
$3 < d < 8 \kpc$.
Its spectral type is in this case
consistent with an early type B3 - A2 V main sequence star.

However, it is interesting to note that the companion star could
be crossing the Hertzsprung gap (U. Kolb, private communication), 
as for GRO J1655-40 \citep{kolb:1997}, although 
the mass of the companion star in the case of V4641 Sgr
is bigger than for GRO J1655-40.
In this view, the location on the CMD would be above
the main sequence, the distance of the object could
then be still bigger than $3 < d < 8 \kpc$, this range
becoming its minimum distance, and the spectral type would be
B3 - A2 IV.
In both possibilities the mass is constrained between 2 $<$ M $<$ 10 $\Msol$,
suggesting that it is an IMXB or a HMXB.
This is consistent with the results derived by \cite{orosz:2000a}.

\subsection{The jets: a new microquasar?}

If the elongation seen in the radio was a moving component, 
the proper motion was between $224 < \mu < 788$ mas/d depending on the
exact time of the ejection.
For the sake of discussion, we will assume that this is the
approaching (brighter) condensation with $\mu_a = 500$ mas/d.
Since $D\leq \frac{c}{\sqrt{\mu_a \mu_r}}$,
and from our results $D \geq 3 \kpc$, we conclude that the apparent
velocity in the plane of the sky would be strongly superluminal,
$v_a$ greater than $8 c$ at the distance of $3 \kpc$.
However, 
no movement of this elongation was detected between Sept. 16.02 and 24.1.
This suggests an interaction with surroundings at 
$0.25 \asec$ $\geq 1.5 \times 10^3$ AU at the distance of 6 kpc.
This is possible if the ejections began to take place at least 10 days before
the radio detection e.g. on September, $8^{th}$, and we can
see from Figure \ref{figure_V-I} that the source was already active
in the optical at this date.

It seems therefore that the activity of this source was not as sporadic
as we could have thought at the beginning.
Indeed, a previous optical outburst 
occured in 1999, August \citep{watanabe:1999}, 
and {\it RXTE} could detect this source during 270 days
before the giant outburst (in't Zand; Markwardt, these proceedings).
Furthermore, during 5 days before the giant outburst, 
the source was in the optical continuously
2 magnitudes brighter than immediately after, showing a modulation
at the orbital period, with no X-ray emission
(typical upper limit of 12 mCrab, \citet{in'tzand:2000}).
All these facts, combined to the high X-ray variability,
show that, although the \citet{orosz:2000a} 
results suggest that the accretion
in the system is of Roche-lobe overflow type, it could exist
in addition a mass loss from the vicinity of the compact
object.
At a distance of $6 \kpc$, the maximum
luminosity of the source is $\sim 4 \times 10^{38} \ergs$, therefore
close to the Eddington limit of a $\sim 10 \Msol$ object 
($1.3 \times 10^{39} \ergs$). If the transfer rate is highly super-Eddington
such a wind could arise.
This wind could be the reason why surrounding matter
was present, allowing the interaction between further ejections
and surrounding matter to take place.
In this case, the companion star is more likely to be a main sequence
star (U. Kolb, private communication).

Finally, it is interesting to note that \citet{marti:2001} observed
this source to look for minute to hour variability, 
discovering a 0.1 mag variability on the time scale of hours.
Among the different interpretations considered by them, 
it is suggested that this variability could originate in an extended corona
surrounding the jets, by analogy with SS 433.
If this interaction between the jets and the surrounding medium is confirmed, 
this source could therefore be added
to the short list of microquasars where such an interaction has
been detected (see e.g. \cite{mirabel:1999} \& \cite{chaty:2001}).

\section{Conclusions}

From our optical and the {\it SAX} observations we constrained
$0.05 < \nh < 0.15 \times 10^{22} \cmmoinsdeux$. From our
optical and NIR colors the distance is $3 < D < 8 \kpc$,
and the companion star would be a main sequence star of spectral type 
B3 - A2 V.
If the source is crossing the Hertzsprung gap,
this determination of the distance would become its minimum distance,
and the spectral type of the companion star would be B3 - A2 IV. 
The system is therefore an IMXB or a HMXB.
From the radio images, the NIR colors, and the optical spectra,
there is a strong suggestion of interaction of the ejecta
of the source with its surroundings.
This surrounding matter could have originated from a
wind created by fluctuations around the central object,
and in this case the companion star would more certainly
be a main sequence star.
Further observations would be useful to confirm this existence
of surrounding matter.

\section*{acknowledgements}
S.C. thanks Rob Hynes for poiting out this new
flaring source on September 1999, Bob Hjellming
for all the spontaneous communications he gave on the radio
observations of this source, and Ulrich Kolb for many stimulating
discussions and careful rereading of the manuscript. 
S.C. is very grateful to the ESO staff for their availability and skills
to perform service observations for override programs,
and in particular to the NTT team: Leonardo Vanzi, Olivier Hainaut,
St\'ephane Brillant and Vanessa Doublier.
S.C. acknowledges
support from grant F/00-180/A from the Leverhulme Trust.
IFM acknowledges partial support from Conicet/Argentina.
JM acknowledges partial support by DGICYT (PB97-0903) and Junta de
Andaluc\'{\i}a (Spain).

   \bibliographystyle{aa}
   \bibliography{science}

%
%
%


\twocolumn
\newpage


\end{document}